\documentclass[10pt,twocolumn,letterpaper]{IEEEtran}
\makeatletter
\def\ps@headings{%
\def\@oddhead{\mbox{}\scriptsize\rightmark \hfil \thepage}%
\def\@evenhead{\scriptsize\thepage \hfil \leftmark\mbox{}}%
\def\@oddfoot{}%
\def\@evenfoot{}}
\makeatother
\IEEEoverridecommandlockouts 
\usepackage{graphicx,latexsym,subfigure,amsmath,epsfig,latexsym,subfigure,amsmath,cite,amssymb}
\usepackage{amsthm,mathrsfs}
\usepackage{graphicx}
\usepackage{epstopdf}
\usepackage{color}
\usepackage{CJK}
\usepackage{cite} 
\usepackage{algorithmic} 
\usepackage{mathrsfs}
\usepackage{bm}
\usepackage{float}
\usepackage{setspace}
\usepackage{color}
\usepackage{subfigure}
\usepackage{balance}
\usepackage{dsfont} 
\usepackage[ruled,linesnumbered]{algorithm2e} 
\usepackage{amsfonts,amssymb}
\usepackage{upgreek} 
\usepackage{geometry}
\SetKwRepeat{DoWhile}{Do}{While}

\begin{document}

\title{\huge Multi-Target Cooperative Visible Light Positioning: A Compressed Sensing Based Framework }

\author{
        \IEEEauthorblockN{Xianyao Wang, and Sicong Liu\IEEEauthorrefmark{1}, \IEEEmembership{Senior Member, IEEE}} \\

\thanks{Xianyao Wang and Sicong Liu are with the Department of Information and Communication Engineering, School of Informatics, Xiamen University, Xiamen 361005, China. Sicong Liu is also with National Mobile Communications Research Laboratory, Southeast University, China. This work is supported in part by the National Natural Science Foundation of China (No. 61901403 and 62077040), in part by the Science and Technology Key Project of Fujian Province, China (No. 2021HZ021004 and 2019HZ020009), in part by the open research fund of National Mobile Communications Research Laboratory, Southeast University (No. 2023D10), and in part by the Youth Innovation Fund of Natural Science Foundation of Xiamen (No.3502Z20206039). (\emph{Corresponding Author: Sicong Liu.} email: liusc@xmu.edu.cn).}
}
\maketitle
\thispagestyle{empty}
\begin{abstract}
In this paper, a compressed sensing (CS) based framework of multi-target cooperative visible light positioning (VLP) is formulated to realize simultaneous high-accuracy localization of multiple targets. The light emitting diodes (LEDs) intended for illumination are utilized to locate multiple target mobile terminals equipped with photodetectors.
{\color{black}The indoor area can be divided into a two-dimensional grid of discrete points, and the targets are located in only a few grid points, which has a sparse property. Thus, the multi-target localization problem can be transferred into a sparse recovery problem. Specifically, a CS-based framework is formulated exploiting the superposition of the received visible light signals at the multiple targets to be located via inter-target cooperation. Then it can be efficiently resolved using CS-based algorithms. Moreover, inter-anchor cooperation is introduced to the CS-based framework by the cross-correlation between the signals corresponding to different LEDs, i.e., anchors, which further improves the localization accuracy.}
Enabled by the proposed CS-based framework and the devised cooperation mechanism, the proposed scheme can simultaneously locate multiple targets with high precision and low computational complexity. Simulation results show that the proposed schemes can achieve centimeter-level multi-target positioning with sub-meter accuracy, which outperforms existing benchmark schemes.
\end{abstract}

\vspace{-0.05in}
\section{Introduction}\label{sec:intro}
{\color{black}Although satellite navigation systems such as Global Positioning System (GPS) has been widely applied in outdoor positioning and navigation services, it is still difficult for them to achieve accurate positioning in tunnels, offices, and other indoor environments due to severe occlusion and attenuation of satellite navigation signals \cite{r1}.
Thus, the existing indoor signals such as radio frequency wireless signals, including WiFi, Bluetooth, and ultra-wide band signals, and also the visible light signals \cite{r2}, can be employed for indoor positioning instead of the satellite navigation signals.
}

For radio frequency wireless positioning, a method commonly applied is to estimate the position from the measurements of received signal strength (RSS) detected by the sensors. An indoor positioning method is proposed based on location fingerprints mapping \cite{r3}, where the space is divided into a set of grid points, and the position of the target is determined by matching the actual RSS measurements and the stored fingerprints data.
However, fingerprint identification based localization usually needs to measure a large amount of information and establish a database, which cost relatively higher complexity and resources \cite{r4}.

In recent years, visible light communication (VLC) technology, which uses the widely deployed light emitting diodes (LEDs) for signal transmission, has drawn great attention because of its advantages of ultra-wide unlicensed spectrum, high energy efficiency, low implementation costs, and anti-electromagnetic interference capability \cite{r5}. Thus, the VLC signal can also be utilized for high-precision indoor positioning, which is called visible light positioning (VLP). Different from radio frequency wireless channels, the visible light channel is mainly dominant by the line-of-sight (LoS) link \cite{r6}, which effectively mitigates the impact of multipath interference and thus improves the positioning accuracy significantly. Existing VLP methods are mainly based on RSS \cite{r7}, time-of-arrival (ToA) \cite{r8} and angle-of-arrival (AoA) \cite{r9}, etc.
{\color{black}However, most of the existing schemes of VLP are aimed at single-target localization. The performance of simultaneous high-precision multi-target VLP still remains to be improved. Moreover, the mutual information between different targets to be located can be combined with a cooperative localization mechanism to achieve a better multi-target joint localization performance.}

If one divides the indoor area into a two-dimensional grid of discrete points, the multiple targets are actually located in only a few grid points, which has a sparse nature. Thus, the multi-target localization problem can be transferred into a sparse recovery problem, which can be solved efficiently by the compressed sensing (CS) based algorithms. For indoor wireless localization tasks using radio frequency signals, H. Jamali-Rad \emph{et al} proposed a CS-based multi-target localization method \cite{r10}, which can locate multiple targets simultaneously with high accuracy. N. Garcia \emph{et al} proposed a direct source localization method in massive MIMO systems, which achieves robust sub-meter localization precision in the presence of multipath fading \cite{r11}. However, wireless signal will undergo a large number of reflections when there are many scatters in the room, resulting in severe multipath interference, which might limit the positioning performance \cite{r12}. 

{\color{black}R. Zhang \emph{et al} introduced the CS technique to multi-target VLP \cite{r13}, where a reverse visible light multi-target localization scheme based on sparse matrix reconstruction was devised. In this scheme, some photodetectors (PDs) mounted on the ceiling are used to detect the uplink visible light signals sent from the target user terminals equipped with LEDs.}
{\color{black} In realistic indoor environments, VLP is mainly implemented using the existing illumination LEDs on the ceiling in a downlink manner. Hence, a different downlink framework of CS-based multi-target VLP still remains to be investigated.}

To resolve the difficulty of the existing schemes, this paper formulates a CS-based framework of multi-target cooperative VLP, which uses only a few measurements of the visible light signals received by the target terminals to achieve simultaneous multi-target high-precision positioning. A CS-based multi-target framework of VLP is formulated to recover the sparse locations of the target terminals in the grid from the received visible light signals. In this way, the multi-target localization problem is transferred into a sparse recovery problem, which is solved efficiently by CS-based algorithms. Inter-target cooperation is utilized to facilitate the formulation of the CS-based framework. Moreover, inter-anchor cooperation is achieved by cross-correlation between the signals corresponding to different LEDs, which is exploited to further improve the localization performance in realistic complex indoor environments.

The rest of this article will be organized as follows: {\color{black} Section II presents the specific system model and briefly explains the basic principles. Section III gives a detailed description of the proposed CS-based framework of multi-target cooperative VLP. Section IV presents the preliminary simulation results. Finally, a conclusion is made in Section V.}
\vspace{-0.10in}
\section{System Model}\label{sec:sm}

\subsection{Visible Light Channel Model}
{\color{black}The LEDs intended for illumination are employed as light source for VLP, which follows the Lambertian radiation pattern \cite{r14}, and the irradiance intensity is expressed as
\begin{equation}\label{sec3:cs}\small
\begin{split}
{R_{\rm{o}}}\left( \alpha  \right) = \frac{{m{\rm{ + 1}}}}{{2\pi }}{\cos ^m}\left( \alpha  \right), - \frac{\pi }{2} \le \alpha  \le \frac{\pi }{2},
\end{split}
\end{equation}
where $\alpha $ is the irradiation angle of the LED; $m$ is the Lambertian radiation ordinal of the LED given by $m = {{ - \ln 2} \mathord{\left/
 {\vphantom {{ - \ln 2} {\ln \left( {\cos {\alpha _{{1 \mathord{\left/
 {\vphantom {1 2}} \right.
 \kern-\nulldelimiterspace} 2}}}} \right)}}} \right.
 \kern-\nulldelimiterspace} {\ln \left( {\cos {\alpha _{{1 \mathord{\left/
 {\vphantom {1 2}} \right.
 \kern-\nulldelimiterspace} 2}}}} \right)}}$ with ${\alpha _{{1 \mathord{\left/
 {\vphantom {1 2}} \right.
 \kern-\nulldelimiterspace} 2}}}$ denoting the intensity of the half-power angle \cite{r15}.

As mentioned in the previous section, non-line-of-sight (NLoS) visible light signals are neglectable compared to the LoS link component which is dominant \cite{r6}. Thus, as shown in Fig. 1, the channel gain of the visible light signal propagation in an indoor environment is given by
\begin{equation}\label{sec3:cs}\small
\begin{split}
h = \frac{1}{{{d^2}}}{R_{\rm{o}}}\left( \alpha  \right){A_{{\rm{eff}}}}\left( \varphi  \right),0 \le \varphi  \le {\varphi _{{\rm{FOV}}}},
\end{split}
\end{equation}
where $d$ is the distance between the LED and the PD; $\varphi $ is the angle of incidence at the PD; ${\varphi _{{\rm{FOV}}}}$ is the field of view (FOV) of the PD; ${A_{{\rm{eff}}}}\left( \varphi  \right)$ is the effective detection area of the PD given by
\begin{equation}\label{sec3:cs}\small
{A_{{\rm{eff}}}}\left( \varphi  \right) = {{\mathop{\rm A}\nolimits} _{\det }}{G_{{\rm{filter}}}}{G_{{\rm{conc}}}}\cos \varphi,
\end{equation}
where ${{\mathop{\rm A}\nolimits} _{\det }}$ is the physical area of the detector; ${G_{{\rm{filter}}}}$ and ${G_{{\rm{conc}}}}$ represent the gain of the optical filter and the gain of the optical concentrator, respectively \cite{r16}.

\begin{figure}[!t]
\begin{center}
\vspace{-0.0cm}  
\setlength{\abovecaptionskip}{-0.10cm}   
\setlength{\belowcaptionskip}{-0.08cm}   
\includegraphics[width=2.6 in]{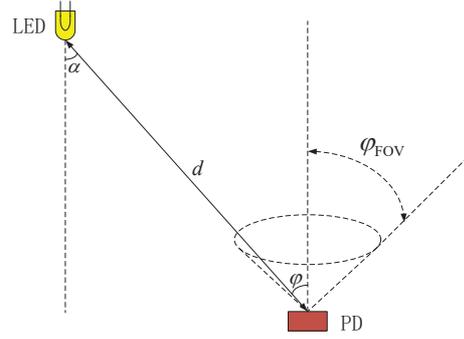}\\
\caption{Channel model of visible light signal propagation for indoor VLP.}
\label{system}
\end{center}
\end{figure}

\subsection{System Model of Indoor Multi-Target Visible Light Positioning}
{\color{black}Utilizing the existing indoor illumination infrastructure, a downlink VLP structure is formulated as shown in Fig. 2. A coordinate system is established in the room for representation of the locations. Each location in this room can be represented by three-dimensional coordinates. On the ceiling, several LEDs are utilized to transmit visible light signals via orthogonal frequency division multiple access (OFDMA), where different LEDs are occupying different orthogonal subcarriers of the VLC spectrum to carry the unique identification and the corresponding location information.

It is assumed that the target mobile terminals are ${d_{\rm{h}}}$ high above the floor, and can be randomly distributed in the two-dimensional plane that is equally divided into $N$ grid points. The center point of each grid is considered as all possible positions of the target terminals. Then, the task to locate the targets is transferred to the problem of finding the grid points the target terminals are located in.

Let $M$ denote the number of LEDs, and the coordinates of the LEDs can be expressed as
\begin{equation}
\left( {x_i^{{\rm{(tx)}}},y_i^{{\rm{(tx)}}},z_i^{{\rm{(tx)}}}} \right),i = 1, \cdots ,M.
\end{equation}
Similarly, the coordinates of the grid points can be expressed as
\begin{equation}
\left( {x_j^{({\rm{rx}})},y_j^{({\rm{rx}})},z_j^{({\rm{rx}})}} \right),j = 1, \cdots ,N.
\end{equation}
For convenience of presentation, let ${{\rm{P}}_j}$ denote the $j$-th grid point. Let ${d_{ij}}$ represent the distance from the $i$-th LED to ${{\rm{P}}_j}$, which is given by
\begin{small}
\begin{equation}
{d_{ij}} = \sqrt[2]{{{{\left( {x_i^{{\rm{(tx)}}} - x_j^{{\rm{(rx)}}}} \right)}^2} + {{\left( {y_i^{{\rm{(tx)}}} - y_j^{{\rm{(rx)}}}} \right)}^2} + {{\left( {z_i^{{\rm{(tx)}}} - z_j^{{\rm{(rx)}}}} \right)}^2}}}.
\end{equation}
\end{small}
Substituting (6) into the equation (2), the channel gain from from the $i$-th LED to ${{\rm{P}}_j}$, i.e., ${h_{ij}}$, can be obtained. Let ${x_i}$ represent the transmit pilot signal sent by the $i$-th LED, which is known at the receiver used for localization. Suppose a target is located at ${{\rm{P}}_j}$, and then the received pilot signal received by the target at ${{\rm{P}}_j}$ sent from the $i$-th LED can be denoted as
\begin{equation}
{y_{ij}} = {h_{ij}}{x_i} + {\omega _{ij}},
\end{equation}
where ${\omega _{ij}}$ represents the background noise of the link between the $i$-th LED and the target at ${{\rm{P}}_j}$. The channel gain ${h_{ij}}$ can be estimated for each link between an LED and a grid point, by performing channel estimation methods using the received pilot signals.

\begin{figure}[!t]
\begin{center}
\vspace{-0.2cm}  
\setlength{\abovecaptionskip}{-0.10cm}   
\setlength{\belowcaptionskip}{-0.08cm}  
\includegraphics[width=2.6 in]{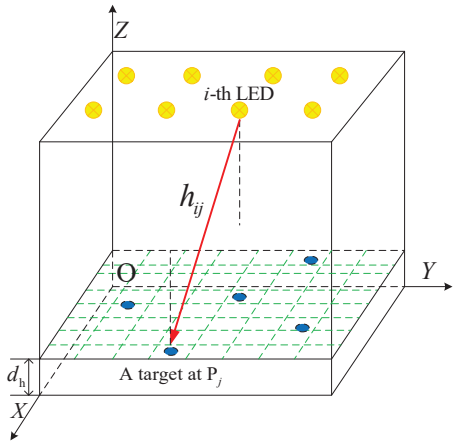}\\
\caption{Geometric architecture of indoor multi-target VLP system: LEDs are utilized as visible light transmitters; Multiple targets equipped with PDs to be located are randomly distributed in the area divided into a set of grid points.}
\label{system} 
\end{center}
\end{figure}

\vspace{-0.10in}
\section{COMPRESSED SENSING BASED FRAMEWORK OF MULTI-TARGET COOPERATIVE VISIBLE LIGHT POSITIONING}\label{sec:method}
\subsection{Compressed Sensing Based Multi-Target Visible Light Positioning via Inter-Target Cooperation}
In order to realize accurate multi-target localization in a typical downlink VLP system, we formulate a framework of CS-based multi-target VLP (CSM-VLP) in this section. Specifically, consider a generalized indoor VLP scenario as illustrated in Fig. 2, and assume that the number of target terminals is $K$, which is far smaller than the number of grid points in the area, i.e., $K \ll N$.  The received pilot signals ${\rm{\{ }}{y_{ik}}{\rm{\} }}_{k = 1}^K$ at all the $K$ target terminals sent from the $i$-th LED can be aggregated via inter-target cooperation such as wireless communication links. Let ${y_i}$ denote the aggregated received signal, i.e., the sum of the $K$ received pilot signals, which is given by
\begin{equation}
{y_i} = \sum\limits_{k = 1}^K {{y_{ik}}}  = \sum\limits_{k = 1}^K {{h_{ik}}} {x_i} + {\omega _i},
\end{equation}
where ${\omega _i}$ represents the sum of the background noise of the transmission links from the $i$-th LED to all the $K$ target terminals. Then, we can formulate an aggregated received signal vector ${\bf{y}}$ by combining all the aggregated received signals ${\rm{\{ }}{y_i}{\rm{\} }}_{i = 1}^M$ corresponding to all the $M$ LEDs, which can be expressed as
\begin{small}
\begin{equation}
{\bf{y}} = \left[ {\begin{array}{*{20}{c}}
{{y_1}}\\
{{y_2}}\\
 \vdots \\
{{y_M}}
\end{array}} \right] = \left[ {\begin{array}{*{20}{c}}
{\sum\limits_{k = 1}^K {{h_{1k}}} {x_1}}\\
{\sum\limits_{k = 1}^K {{h_{2k}}} {x_2}}\\
 \vdots \\
{\sum\limits_{k = 1}^K {{h_{Mk}}} {x_M}}
\end{array}} \right] + {{\bm{\upomega }}},
\end{equation}
\end{small}
where ${\bm{\upomega }}$ is the aggregated background noise vector.
And then a CS-based framework of multi-target VLP can be formulated based on (9) exploiting the inherent sparsity of the target locations with respect to the overall grid points in the indoor area, which is given by
\begin{equation}
{\bf{y}} = \left[ {\begin{array}{*{20}{c}}
{{y_1}}\\
{{y_2}}\\
 \vdots \\
{{y_M}}
\end{array}} \right] = \underbrace {\left[ {\begin{array}{*{20}{c}}
{{h_{11}}{x_1}}& \cdots &{{h_{1N}}{x_1}}\\
{{h_{21}}{x_2}}& \cdots &{{h_{2N}}{x_2}}\\
 \vdots & \ddots & \vdots \\
{{h_{M1}}{x_M}}& \cdots &{{h_{MN}}{x_M}}
\end{array}} \right]}_{\bm{\Phi }}{\bm{\uptheta }} + {\bm{\upomega }},
\end{equation}
where the on-grid target localization vector ${\bm{\uptheta }} = {\left[ {{\theta _1}, \cdots ,{\theta _j}, \cdots ,{\theta _N}} \right]^{\rm{T}}}$ is an indicator vector with $N$ elements, with each element corresponding to a grid point in the area, and the value of each element indicates whether there is a target located in the corresponding grid point or not. Thus, most of the elements are mostly zeros except that the elements corresponding to the positions of the $K$ targets to be located are ones. Since the number of the nonzero elements in vector ${\bm{\uptheta }}$ is much smaller than the length of the vector, it is a sparse vector that can be recovered through sparse recovery methods. The observation matrix ${\bm{\Phi }}$ in the CS-based measurement model (10) is defined as
\begin{small}
\begin{equation}
{\bf{\Phi }} = \left[ {\begin{array}{*{20}{c}}
{{h_{11}}{x_1}}& \cdots &{{h_{1N}}{x_1}}\\
{{h_{21}}{x_2}}& \cdots &{{h_{2N}}{x_2}}\\
 \vdots & \ddots & \vdots \\
{{h_{M1}}{x_M}}& \cdots &{{h_{MN}}{x_M}}
\end{array}} \right].
\end{equation}
\end{small}

In order to refine the CS-based model in (10) to make it a more universal framework independent of different pilot signals, we can get the power of the aggregated received signals via autocorrelation of the aggregated received signal vector ${\bf{y}}$\cite{r17}, which yields the received power ${{\bf{p}}_{{\rm{rx}}}}$ as represented by
\begin{equation}
\begin{aligned}
\mathbf{{{\bf{p}}_{{\rm{rx}}}}} &=\mathbb{E}\left\{\mathbf{y} \odot \mathbf{y}^*\right\} \\
&=\mathbb{E}\left\{(\boldsymbol{\Phi} \boldsymbol{\uptheta}+\boldsymbol{\upomega}) \odot(\boldsymbol{\Phi} \boldsymbol{\uptheta}+\boldsymbol{\upomega})^*\right\} \\
&=\mathbb{E}\left\{\boldsymbol{\Phi} \odot \boldsymbol{\Phi}^*\right\} \boldsymbol{\uptheta}+\mathbb{E}\left\{\boldsymbol{\upomega} \odot \boldsymbol{\upomega}^*\right\} \\
&=\mathbf{J} \boldsymbol{\uptheta}+\sigma_n^2 \mathbf{1}_M,
\end{aligned}
\end{equation}
where $\mathbb{E}\left\{  \cdot  \right\}$ represents the statistical expectation operator; $\odot$ denotes the Hadamard product operator; ${\left(  \cdot  \right)^ * }$ represents the complex-conjugate operator; $\sigma _n^2$ denotes the variance of the background noise; ${{\bf{1}}_M}$ is an all-one-valued $M$-length vector. Then a refined CS-based power measurement model is established in equation (12), where the observation matrix ${\bf{J}}$ can also be regarded as a fingerprint database of the indoor visible light channels and propagation environments used for localization, which is given by
\begin{small}
\begin{equation}
{\bf{J}} = \left[ {\begin{array}{*{20}{c}}
{{{\left| {{h_{11}}} \right|}^2}}&{{{\left| {{h_{12}}} \right|}^2}}& \cdots &{{{\left| {{h_{1N}}} \right|}^2}}\\
{{{\left| {{h_{21}}} \right|}^2}}&{{{\left| {{h_{22}}} \right|}^2}}& \cdots &{{{\left| {{h_{2N}}} \right|}^2}}\\
 \vdots & \vdots & \ddots & \vdots \\
{{{\left| {{h_{M1}}} \right|}^2}}&{{{\left| {{h_{M2}}} \right|}^2}}& \cdots &{{{\left| {{h_{MN}}} \right|}^2}}
\end{array}} \right].
\end{equation}
\end{small}
It is observed that the elements in ${\bf{J}}$ are related to the channel gains $\left\{ {{h_{ij}}} \right\}$. During the positioning phase, the refined CS-based power measurement model is obtained via the downlink VLP system with inter-target cooperation. It can be found from (12) that this is actually a sparse recovery problem aimed to recover the on-grid target localization sparse vector ${\bm{\uptheta }}$ through the power measurements ${{\bf{p}}_{{\rm{rx}}}}$ with the given observation matrix ${\bf{J}}$. The grid points corresponding to the positions of the nonzero elements in ${\bm{\uptheta }}$ are the estimated positions of the target terminals. The sparse recovery problem can be efficiently solved using ${\ell _1}$-norm minimization methods or CS-based greedy algorithms such as orthogonal matching pursuit (OMP).

In fact, the accuracy of the proposed scheme is related to the step-size of the grid point. A denser grid partition makes the quantization error of the location coordinates smaller. Appropriately increasing the number of grid points can improve the positioning resolution. Besides, the increase of $N$ also makes ${\bm{\uptheta }}$ a sparser vector, which is easier to be recovered by CS-based algorithms. However, it is not necessarily true that a smaller grid step-size will lead to better CS-based localization performance. According to the CS theory, the sparse recovery problem as established in (12) can be solved effectively if the condition $M \ge {\rm{\upmu }}K\log \left( {{N \mathord{\left/
 {\vphantom {N K}} \right.
 \kern-\nulldelimiterspace} K}} \right)$ is satisfied, where ${\rm{\upmu }}$ is a positive constant \cite{r18}. Thus, it is necessary to carefully choose $M$ and $N$ according to the actual situation.

\subsection{Enhanced Cooperative Compressed Sensing Based Multi-Target Visible Light Positioning via Inter-Anchor Cooperation}
Although the CSM-VLP scheme has exploited the inter-target information to formulate the aggregated measurement data as given in (10), the potential geometric information of the indoor environment and the channel information corresponding to different anchors, i.e., LEDs, have not been fully utilized to enhance the localization robustness. Thus, we further devise a cooperative CSM-VLP (CoCSM-VLP) scheme via the cross-correlation between the aggregated received signals corresponding to different LEDs, in order to make use of the inter-anchor cooperation and achieve a better multi-target joint localization performance.
Specifically, we obtain the inter-anchor correlation ${{\bf{P}}_{{\rm{corr}}}}$ via the cross-correlation between the aggregated received signals as provided in (10), which is given by
\begin{equation}
\begin{aligned}
\mathbf{P}_{\text {corr }} &=\mathbb{E}\left\{\mathbf{y y}^{\mathrm{H}}\right\} \\
&=\mathbb{E}\left\{(\boldsymbol{\Phi} \boldsymbol{\uptheta}+\boldsymbol{\upomega})(\boldsymbol{\Phi} \boldsymbol{\uptheta}+\boldsymbol{\upomega})^{\mathrm{H}}\right\} \\
&=\mathbb{E}\left\{\boldsymbol{\Phi} \boldsymbol{\uptheta} \boldsymbol{\uptheta}^{\mathrm{H}} \boldsymbol{\Phi}^{\mathrm{H}}\right\}+\mathbb{E}\left\{\boldsymbol{\upomega} \boldsymbol{\upomega}^{\mathrm{H}}\right\},
\end{aligned}
\end{equation}
where ${\left(  \cdot  \right)^{\rm{H}}}$ represents the conjugate-transpose operator. Then, we can vectorize both side of (14) to derive the cross-correlation measurement vector as given by
\begin{equation}
\begin{aligned}
\mathbf{p}_{\text {corr }} &=\operatorname{vec}\left(\mathbf{P}_{\text {corr }}\right) \\
&=\operatorname{vec}\left(\mathbb{E}\left\{\boldsymbol{\Phi} \boldsymbol{\uptheta} \boldsymbol{\uptheta}^{\mathrm{H}} \boldsymbol{\Phi}^{\mathrm{H}}\right\}\right)+\operatorname{vec}\left(\mathbb{E}\left\{\boldsymbol{\Phi} \boldsymbol{\uptheta} \boldsymbol{\uptheta}^{\mathrm{H}} \boldsymbol{\Phi}^{\mathrm{H}}\right\}\right) \\
&=\mathbb{E}\left\{\boldsymbol{\Phi}^* \otimes \boldsymbol{\Phi}\right\} \operatorname{vec}\left(\boldsymbol{\uptheta} \boldsymbol{\uptheta}^{\mathrm{H}}\right)+\operatorname{vec}\left(\boldsymbol{\upomega} \boldsymbol{\upomega}^{\mathrm{H}}\right) \\
&=\mathbb{E}\left\{\boldsymbol{\Phi}^* \cdot \boldsymbol{\Phi}\right\} \boldsymbol{\uptheta}+\operatorname{vec}\left(\sigma_n^2 \mathbf{I}_M\right),
\end{aligned}
\end{equation}
where $ \otimes $ represents the Kronecker product operator; $\cdot$ denotes Khatri-Rao product; ${{\bm{{\rm I}}}_M}$ is a unit matrix of size $M \times M$; the cross-correlation measurement vector ${{\bf{p}}_{{\rm{corr}}}}$ is of length-${M^2}$.

In fact, the problem given in (15) is a system of linear equations, of which the number of independent equations has a direct influence on the CS-based sparse recovery performance. This is actually determined by the coherence property of the observation matrix based on the CS theory. Since the cross-correlation matrix ${{\bf{P}}_{{\rm{corr}}}}$ in (14) is symmetrical, we can define an selection matrix ${\bf{S}}$ of size ${{M\left( {M + 1} \right)} \mathord{\left/
 {\vphantom {{M\left( {M + 1} \right)} 2}} \right.
 \kern-\nulldelimiterspace} 2} \times {M^2}$ to select the independent equations of (15), which is corresponding to the rows corresponding to the $M$ diagonal and ${{M\left( {M - 1} \right)} \mathord{\left/
 {\vphantom {{M\left( {M - 1} \right)} 2}} \right.
 \kern-\nulldelimiterspace} 2}$ upper-diagonal elements of ${{\bf{P}}_{{\rm{corr}}}}$. After the selection process, we can formulate the CS-based cross-correlation measurement model as given by
\begin{equation}
\begin{aligned}
\hat{\mathbf{y}} &={\bf{S}} \operatorname{vec}\left(\mathbf{P}_{\text {corr }}\right) \\
&={\bf{S}} \mathbb{E}\left\{\boldsymbol{\Phi}^* \cdot \boldsymbol{\Phi}\right\} \boldsymbol{\uptheta}+{\bf{S}} \operatorname{vec}\left(\sigma_n^2 \mathbf{I}_M\right) \\
&=\boldsymbol{\Psi} \boldsymbol{\uptheta}+\boldsymbol{\upomega}_{\mathrm{vec}},
\end{aligned}
\end{equation}
where the size ${{M\left( {M + 1} \right)} \mathord{\left/
 {\vphantom {{M\left( {M + 1} \right)} 2}} \right.
 \kern-\nulldelimiterspace} 2} \times N$ matrix ${\bf{\Psi }}$ represents the observation matrix, i.e., the fingerprint database of the CoCSM-VLP scheme, for the formulated CS-based cross-correlation measurement model in (16). Compared with the observation matrix ${\bf{J}}$ in (12), the number of rows in ${\bf{\Psi }}$ is significantly higher, which means that more independent equations are available to solve for the support, i.e., positions of the nonzero elements of the unknown sparse localization vector ${\bm{\uptheta }}$. All the elements in ${\bf{\Psi }}$ are also related to the channel gains $\left\{ {{h_{ij}}} \right\}$, and thus the channel information required to construct ${\bf{\Psi }}$ can also be collected via beforehand channel estimation indoors, which is given by
\begin{small}
\begin{equation}
{\bf{\Psi }} = \left[ {\begin{array}{*{20}{c}}
{{{\left| {{h_{11}}} \right|}^2}}&{{{\left| {{h_{12}}} \right|}^2}}& \cdots &{{{\left| {{h_{1N}}} \right|}^2}}\\
{h_{11}^ * {h_{21}}}&{h_{12}^ * {h_{22}}}& \cdots &{h_{1N}^ * {h_{2N}}}\\
{{{\left| {{h_{21}}} \right|}^2}}&{{{\left| {{h_{22}}} \right|}^2}}& \cdots &{{{\left| {{h_{2N}}} \right|}^2}}\\
 \vdots & \vdots & \vdots & \vdots \\
{h_{11}^ * {h_{M1}}}&{h_{12}^ * {h_{M2}}}& \cdots &{h_{1N}^ * {h_{MN}}}\\
 \vdots & \vdots & \vdots & \vdots \\
{{{\left| {{h_{M1}}} \right|}^2}}&{{{\left| {{h_{M2}}} \right|}^2}}& \cdots &{{{\left| {{h_{MN}}} \right|}^2}}
\end{array}} \right].
\end{equation}
\end{small}

Through the inter-anchor cooperation procedures, the number of available independent linear equations in the CS-based cross-correlation measurement model (16) has increased from $M$ to ${{M\left( {M + 1} \right)} \mathord{\left/
 {\vphantom {{M\left( {M + 1} \right)} 2}} \right.
 \kern-\nulldelimiterspace} 2}$ compared with the measurement model in (12). Thus, the amount of available measurement data for CS-based sparse recovery is greatly increased. This will lead to a direct benefit for the CS-based multi-target localization. Since the amount of measurement data in the cross-correlation measurement model (16), i.e., ${{M\left( {M + 1} \right)} \mathord{\left/
 {\vphantom {{M\left( {M + 1} \right)} 2}} \right.
 \kern-\nulldelimiterspace} 2}$, is much larger than $M$, it is much easier to satisfy the requirement. This will improve the accuracy and reliability of the proposed scheme, especially in severe conditions such as large number of unknown targets ($K$ is large), a dense grid network ($N$ is large), and intensive background noise (the SNR is low).

Afterwards, the sparse recovery problem as given in (16) can be efficiently solved to reconstruct the sparse localization vector ${\bm{\uptheta }}$ similarly as in CSM-VLP, using ${\ell _1}$-norm minimization methods or CS-based greedy algorithms such as orthogonal matching pursuit (OMP).
\vspace{-0.10in}
\section{SIMULATION RESULTS AND DISCUSSIONS}\label{sec:result}
In this section, the performance of the proposed CSM-VLP and CoSM-VLP schemes is evaluated through simulation experiments, which is conducted in a typical geometric architecture of indoor multi-target VLP as shown in Fig. 2. The size of the room is ${\rm{4}} \times {\rm{4}} \times {\rm{3}}$ ${{\rm{m}}^3}$. According to the system model presented in Section II, the plane is divided into $400$ grid points with an identical size of  ${\rm{0}}{\rm{.2}} \times {\rm{0}}{\rm{.2}}$ ${{\rm{m}}^2}$. Without loss of generality, 16 evenly spaced LED light fixtures placed at the ceiling used for illumination are employed to transmit the VLP signals.

The localization performance of the two proposed CSM-VLP and CoCSM-VLP schemes is reported in Fig. 3, where the estimated and the ground-truth locations are all depicted on the plane for comparison, and the localization error is indicated by a short line connecting them. It can be observed from Fig. 3 that the proposed CSM-VLP scheme has a relatively accurate localization performance for the eight target terminals with unknown random positions. It is also noted that the localization performance is limited for a few of the targets. We can observe from Fig. 3 that the multi-target localization performance is further improved using the proposed enhanced CoCSM-VLP scheme compared to the CSM-VLP scheme. It is shown that the additional information among the localization signals corresponding to different LEDs is made full use of via inter-anchor cooperation to greatly improve the multi-target localization accuracy.

\begin{figure}[!t]
\begin{center}
\vspace{-0.0cm}  
\setlength{\abovecaptionskip}{-0.10cm}   
\setlength{\belowcaptionskip}{-0.08cm}  
\includegraphics[width=2.8 in]{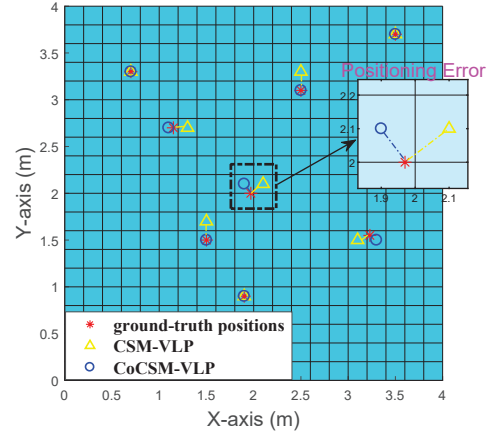}\\
\caption{Localization performance of the two proposed CSM-VLP and CoCSM-VLP schemes for simultaneous multi-target for eight target terminals with unknown random locations in the indoor plane; The ground-truth positions are marked by red dots in this figure.}
\label{system} 
\end{center}
\end{figure}

Next, we calculate the Euclidean distance between the locations estimated by the two proposed schemes and the ground-truth locations of the unknown multiple target terminals to measure the positioning error. The average positioning error $\Delta $ reflecting the positioning accuracy is given by
\begin{equation}
\Delta  = \frac{1}{K}\sum\limits_1^K {\sqrt[2]{{{{\left( {x_k^{({\rm{rx}})} - x_k^{({\rm{real}})}} \right)}^2} + {{\left( {y_k^{({\rm{rx}})} - y_k^{({\rm{real}})}} \right)}^2}}}},
\end{equation}
where $(x_k^{({\rm{real}})},y_k^{({\rm{real}})}),k = 1, \cdots ,K$ denote the ground-truth coordinates of the target terminals.

As reported in Fig. 4, the performance of the proposed CSM-VLP and CoCSM-VLP schemes with respect to the number of target terminals is investigated. A conventional RSS-based VLP method \cite{r19} is also evaluated for comparison. As can be seen from Fig. 4, both the two proposed CS-based schemes of CSM-VLP and CoCSM-VLP outperform the conventional scheme significantly, which verifies the effectiveness of the formulated CS-based framework of multi-target VLP. It is also observed that the enhanced CoCSM-VLP scheme further outperforms the CSM-VLP scheme, which validates the superiority of the inter-anchor cooperative mechanism of the CoCSM-VLP scheme. Meanwhile, it is shown that the average positioning error of both the two proposed schemes and the conventional method all increases with the number of targets to be located. However, it is shown by Fig. 4 that the enhanced CoCSM-VLP scheme is less sensitive to the increase of the number of targets compared to the CSM-VLP scheme. This is because the amount of measurement data in the cross-correlation measurement model (16) is much greater than that of the available measurement data in (12), and more available measurement data leads to a better performance of CS-based sparse recovery, especially in severe conditions such as large number of unknown targets.

\begin{figure}[!t]
\begin{center}
\vspace{-0.0cm}  
\setlength{\abovecaptionskip}{-0.10cm}   
\setlength{\belowcaptionskip}{-0.08cm}  
\includegraphics[width=2.6 in]{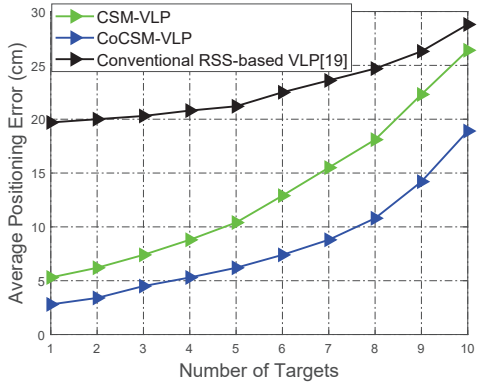}\\
\caption{Average positioning error of the two proposed CSM-VLP and CoCSM-VLP schemes with respect to the number of unknown targets to be located.}
\label{system} 
\end{center}
\end{figure}

\section{CONCLUSION}
In this paper, a CS-based multi-target cooperative framework of indoor visible light positioning (VLP) has been formulated to realize accurate multi-target localization in a typical downlink indoor VLP system employing the LEDs intended for illumination. By fully exploiting the inherent sparsity of the locations of a few targets with respect to all the grid points of the indoor environment, the multi-target localization problem has been transferred into a sparse recovery problem. Thus, a scheme of CS-based multi-target VLP (CSM-VLP) is devised, in which a sparse unknown vector for on-grid target localization is reconstructed from the aggregated received VLP signals as measurement data obtained through inter-target cooperation. Moreover, an enhanced cooperative CSM-VLP scheme (CoCSM-VLP) is devised, which implements inter-anchor cooperation via the cross-correlation between the VLP signals corresponding to different LEDs. Through the two-dimensional cooperation mechanism in the proposed framework of multi-target VLP, the geometric information of the indoor environment and the channel state information can be fully utilized to further improve the positioning performance. The simulation results have shown that the proposed schemes can realize large number multi-target VLP with high accuracy and efficiency in complex indoor environments.

\bibliography{CS_VLP_ICC}

\begin{thebibliography}{10}

\bibitem{r1}
D.~Dardari, P.~Closas, and P.~M. Djurić, ``Indoor tracking: Theory, methods,
  and technologies,'' {\em IEEE Transactions on Vehicular Technology}, vol.~64,
  no.~4, pp.~1263--1278, Apr. 2015.

\bibitem{r2}
T.~Wei, S.~Liu, and X.~Du, ``Visible light integrated positioning and
  communication: A multi-task federated learning framework,'' {\em IEEE
  Transactions on Mobile Computing}, pp.~1--18, 2022.

\bibitem{r3}
C.~Feng, W.~S.~A. Au, S.~Valaee, and Z.~Tan, ``Received-signal-strength-based
  indoor positioning using compressive sensing,'' {\em IEEE Transactions on
  Mobile Computing}, vol.~11, no.~12, pp.~1983--1993, 2012.

\bibitem{r4}
A.~Khalajmehrabadi, N.~Gatsis, and D.~Akopian, ``Modern {WLAN} fingerprinting
  indoor positioning methods and deployment challenges,'' {\em IEEE
  Communications Surveys $\&$ Tutorials}, vol.~19, no.~3, pp.~1974--2002, 2017.

\bibitem{r5}
L.~Xiao, G.~Sheng, S.~Liu, H.~Dai, M.~Peng, and J.~Song, ``Deep reinforcement
  learning-enabled secure visible light communication against eavesdropping,''
  {\em IEEE Transactions on Communications}, vol.~67, no.~10, pp.~6994--7005,
  2019.

\bibitem{r6}
Z.~Ghassemlooy, S.~Arnon, M.~Uysal, Z.~Xu, and J.~Cheng, ``Emerging optical
  wireless communications-advances and challenges,'' {\em IEEE Journal on
  Selected Areas in Communications}, vol.~33, no.~9, pp.~1738--1749, 2015.

\bibitem{r7}
D.~Su, X.~Liu, and S.~Liu, ``Three-dimensional indoor visible light
  localization: A learning-based approach,'' {\em ACM UbiComp'21},
  pp.~672--677, Sept.2021.

\bibitem{r8}
T.~Q. Wang, Y.~A. Sekercioglu, A.~Neild, and J.~Armstrong, ``Position accuracy
  of time-of-arrival based ranging using visible light with application in
  indoor localization systems,'' {\em Journal of Lightwave Technology},
  vol.~31, no.~20, pp.~3302--3308, 2013.

\bibitem{r9}
B.~Zhu, J.~Cheng, Y.~Wang, J.~Yan, and J.~Wang, ``Three-dimensional vlc
  positioning based on angle difference of arrival with arbitrary tilting angle
  of receiver,'' {\em IEEE Journal on Selected Areas in Communications},
  vol.~36, no.~1, pp.~8--22, 2018.

\bibitem{r10}
H.~Jamali-Rad, H.~Ramezani, and G.~Leus, ``Sparse multi-target localization
  using cooperative access points,'' in {\em 2012 IEEE 7th Sensor Array and
  Multichannel Signal Processing Workshop (SAM)}, pp.~353--356, 2012.

\bibitem{r11}
N.~Garcia, H.~Wymeersch, E.~G. Larsson, A.~M. Haimovich, and M.~Coulon,
  ``Direct localization for massive {MIMO},'' {\em IEEE Transactions on Signal
  Processing}, vol.~65, no.~10, pp.~2475--2487, 2017.

\bibitem{r12}
W.~Zhao, S.~Han, W.~Meng, D.~Sun, and R.~Q. Hu, ``{BSDP}: Big sensor data
  preprocessing in multi-source fusion positioning system using compressive
  sensing,'' {\em IEEE Transactions on Vehicular Technology}, vol.~68, no.~9,
  pp.~8866--8880, 2019.

\bibitem{r13}
R.~Zhang, W.-D. Zhong, K.~Qian, S.~Zhang, and P.~Du, ``A reversed visible light
  multitarget localization system via sparse matrix reconstruction,'' {\em IEEE
  Internet of Things Journal}, vol.~5, no.~5, pp.~4223--4230, 2018.

\bibitem{r14}
L.~E.~M. Matheus, A.~B. Vieira, L.~F.~M. Vieira, M.~A.~M. Vieira, and
  O.~Gnawali, ``Visible light communication: Concepts, applications and
  challenges,'' {\em IEEE Communications Surveys $\&$ Tutorials}, vol.~21,
  no.~4, pp.~3204--3237, 2019.

\bibitem{r15}
D.~Karunatilaka, F.~Zafar, V.~Kalavally, and R.~Parthiban, ``{LED} based indoor
  visible light communications: State of the art,'' {\em IEEE Communications
  Surveys $\&$ Tutorials}, vol.~17, no.~3, pp.~1649--1678, 2015.

\bibitem{r16}
Y.~Wang, M.~Chen, Z.~Yang, T.~Luo, and W.~Saad, ``Deep learning for optimal
  deployment of {UAVs} with visible light communications,'' {\em IEEE
  Transactions on Wireless Communications}, vol.~19, no.~11, pp.~7049--7063,
  2020.

\bibitem{r17}
D.~Li, B.~Zhang, and C.~Li, ``A feature-scaling-based $k$-nearest neighbor
  algorithm for indoor positioning systems,'' {\em IEEE Internet of Things
  Journal}, vol.~3, no.~4, pp.~590--597, 2016.

\bibitem{r18}
J.~W. Choi, B.~Shim, Y.~Ding, B.~Rao, and D.~I. Kim, ``Compressed sensing for
  wireless communications: Useful tips and tricks,'' {\em IEEE Communications
  Surveys $\&$ Tutorials}, vol.~19, no.~3, pp.~1527--1550, 2017.

\bibitem{r19}
N.~Huang, C.~Gong, J.~Luo, and Z.~Xu, ``Design and demonstration of robust
  visible light positioning based on received signal strength,'' {\em Journal
  of Lightwave Technology}, vol.~38, no.~20, pp.~5695--5707, 2020.

\end{thebibliography}
\bibliographystyle{IEEEtr}
\end{document}